# Universality Class of $O(N)$ Models

**Adrian Patrascioiu**

Physics Department, University of Arizona
Tucson, AZ 85721, U.S.A.

and

**Erhard Seiler**

Max-Planck-Institut für Physik
– Werner-Heisenberg-Institut –
Föhringer Ring 6, 80805 Munich, Germany

**Abstract**

We point out that existing numerical data on the correlation length and magnetic susceptibility suggest that the two dimensional $O(3)$ model with standard action has critical exponent $\eta = 1/4$, which is inconsistent with asymptotic freedom. This value of $\eta$ is also different from the one of the Wess-Zumino-Novikov-Witten model that is supposed to correspond to the $O(3)$ model at $\theta = \pi$.

Ever since the seminal paper of Kosterlitz and Thouless [1] it has been accepted that in two dimensions (2D) there is a fundamental difference between the Abelian $O(2)$ and the non-Abelian $O(3)$ models: while the first one undergoes a transition (KT) from a phase with exponential decay at high temperature $T$ ($T = 1/\beta$) to a massless one at low temperature, the latter is in the high temperature phase for any $\beta < \infty$. While the first conjecture was proven rigorously by Fröhlich and Spencer [2], the second one remains unproven.

We have advanced several arguments [3, 4, 5] against this accepted belief and in our opinion *all* $O(N)$ models undergo a KT-like transition at some finite $\beta$. Since however our arguments do not constitute a rigorous demonstration, it may be useful to get a glimpse at what the truth may be via existing numerical data. Our observation pertains to the value of the critical exponent $\eta$, defined as follows: let $\xi$ be the correlation length and $\chi$ the magnetic susceptibility. Then $\eta$ can be defined by the following asymptotic statement:

$$\chi \propto \xi^{2-\eta} \qquad (1)$$

i.e. $\ln \chi - (2 - \eta) \ln \xi$ should go to a constant as $\xi \to \infty$.

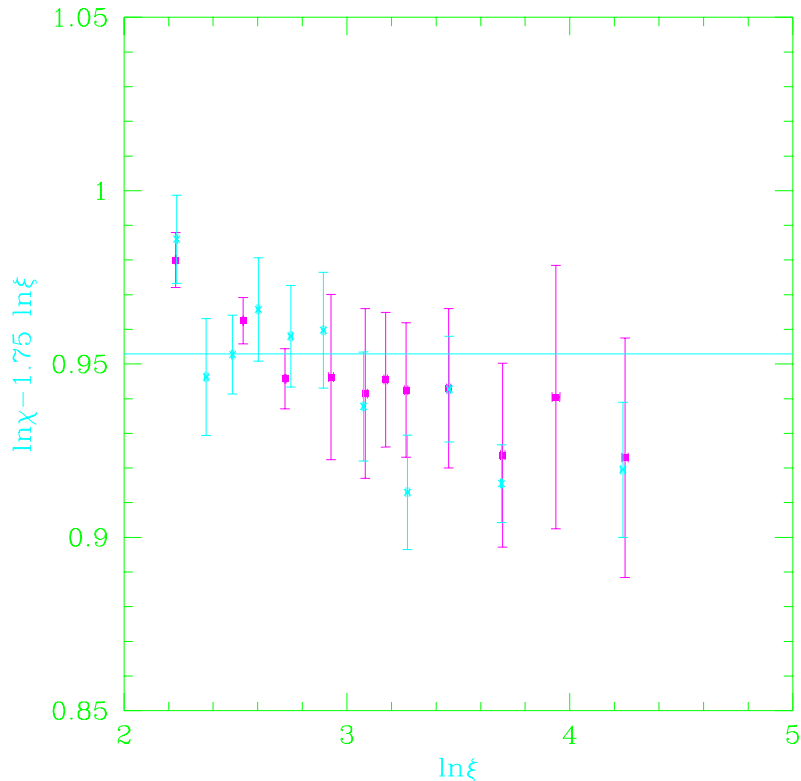

Figure 1: $O(2)$ model, test of $\eta = 1/4$

For the $O(2)$ model certain non-rigorous renormalization group arguments advanced by Kosterlitz [6] predict that $\eta = 1/4$. In Tab.1 we collect the data from [7] and [8]. We could plot $\ln \chi$ vs. $\ln \xi$ and would see a nice straight line with a slope close to 1.75. In fact a fit of $\ln \chi$ to a linear function of $\ln \xi$ yields a slope of $2 - \eta = 1.72$ with a good fit quality, if the errors on both $\xi$ and $\chi$ are taken into account. It is, however, much more significant to look at the deviations from the predicted slope by plotting $\ln \chi - 1.75 \ln \xi$ vs. $\ln \xi$, together with the least square fit to a constant. This is done in Fig.1; the data of [7] are represented by crosses, the ones of [8] by squares. It can be seen that the data still show a tendency to decrease, corresponding to the fact that the aymptotic region has not yet been reached and the fits published in the literature always produced an $\eta > 1/4$, as did ours. But from Fig.1 one can also see a tendency of the data to flatten with increasing $\xi$, in accordance with the theoretical prediction of $\eta = 1/4$. The error bars shown in Fig.1 correspond to adding the errors on $\ln \chi$ to 1.75 times the errors on $\ln \chi$ (probably an overestimation as the errors on $\chi$ and $\xi$ are correlated).

According to the accepted scenario, the behavior of the $O(3)$ model should be vastly different; indeed perturbative renormalization group arguments predict the following asymptotic behavior ("asymptotic scaling"):

$$\xi \propto \frac{\exp(2\pi\beta)}{\beta} \tag{2}$$

$$\chi \propto \frac{\exp(4\pi\beta)}{\beta^4} \tag{3}$$

From this it follows that for $\xi \to \infty$ one should observe $\ln \chi - 2 \ln \xi + 2 \ln \beta$ tending to a constant. To test this prediction, we use the published data of [9] as well as the unpublished ones due to [10], which are collected in Tab.2. In Fig.2 we plot $\ln \chi - 2 \ln \xi + 2 \ln \beta$, which should be constant according to the asymptotic scaling prediction. In Fig.3 we plot instead $\ln \chi - 1.75 \ln \xi$ vs. $\ln \xi$, as we did for $O(2)$. Fig.3 also contains a least square fit to a constant. The data from [9] are represented by squares, the ones of [10] by crosses. The error bars are obtained, as in the case of $O(3)$, by adding the erros on $\ln \chi$ and 1.75 times the errors on $\ln \xi$. Note that in both figures we are using the same scale as in Fig.1. Fitting $\ln \chi$ to a linear function of $\ln \xi$, as we did for $O(2)$, gives a slope of $2 - \eta = 1.74$, again with a good fit quality ($\chi^2 \approx 1$ per degree of freedom).

The following facts are visible from Figs 3 and 2:

(1) The data are not consistent with asymptotic scaling inasmuch the points shown in Fig.2 are not consistent with a constant but show a strong decrease.

(2) The data are consistent with a critical behavior with the same $\eta = 1/4$ that was predicted by the KT theory for the $O(2)$ model.

So we have to conclude that in fact the numerical data disagree with the predictions of asymptotic freedom and instead suggest that both $O(2)$ and $O(3)$ have the same value of $\eta$. The data also indicate that the $O(3)$ model with standard action $\sum s(i) \cdot s(j)$ is not in the same universality class as the $\theta = \pi$ $O(3)$ model, which is supposed to have the Wess-Zumino-Novikov-Witten model as its scaling limit [11]; conformal field theory arguments predict $\eta=1$, a predicition which a recent paper [12] claims to have verified numerically.

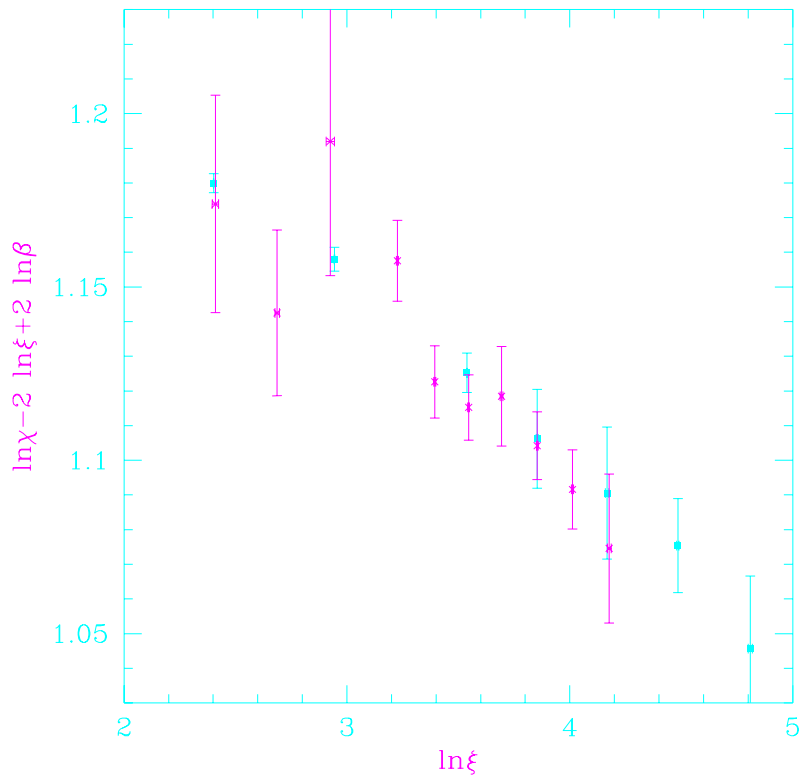

Figure 2: $O(3)$ model, test of asymptotic scaling

A note of caution: in principle $\eta$ can also be determined by studying the finite size scaling of the susceptibility $\chi(L)$. Ideally this study should be performed at $\beta_{crt}$; in practice though, it suffices to place oneself in a regime where $\xi \gg L$. One could ask then what value of $\eta$ would come out of such a determination. The answer is that one must be careful and use a large enough $L$. Indeed, in this type of problems, besides the correlation length $\xi$, there is a second important length $\xi_{PT}$, the distance over which the system is well ordered. From perturbation theory (PT) one knows that for the $O(3)$ model this second length is $O(\exp(\pi\beta))$. Consequently, if one uses periodic boundary conditions (b.c.) and an $L$ and $\beta$ such that $\xi_{PT} \gg L$, then the system will behave according to PT and one will find the perturbative value of $\eta$ which is $1/\pi\beta + O(\beta^{-2})$. As we explained in a recent paper [14], a clear indication that such a determination cannot be trusted to reflect the true thermodynamic value of $\eta$ is the dependence of $\eta$ upon the b.c. employed. Indeed, as we discussed in ref.[14], in the limit $L \to \infty$, the value of $\eta$ must be the same if we use periodic b.c. and freeze two spins situated a distance $L/2$ apart in any arbitrary positions; this then gives a criterion for deciding whether one has used a sufficiently large $L$ and whether the determination of $\eta$ is trustworthy.

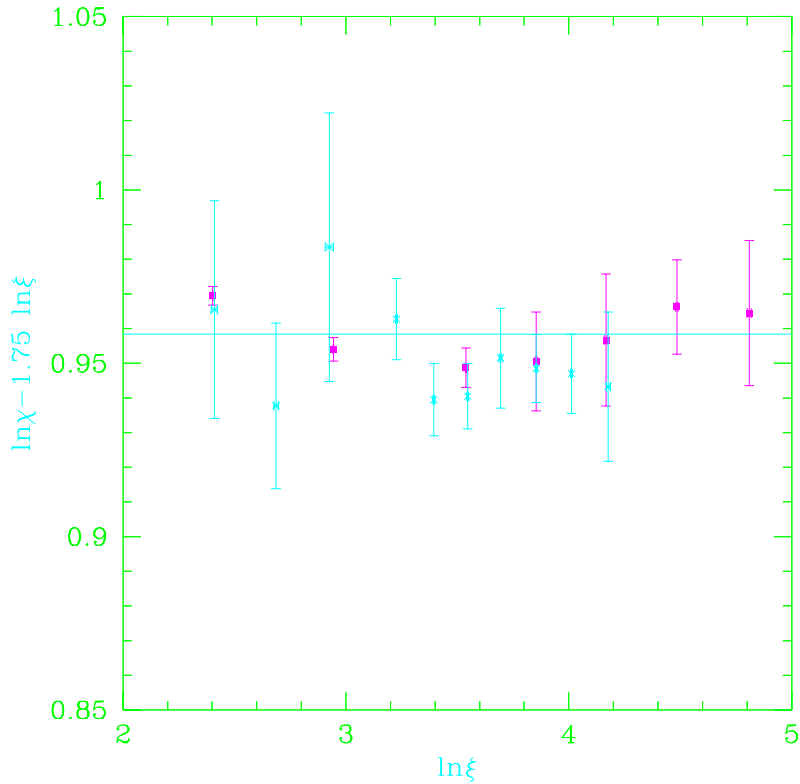

Figure 3: $O(3)$ model, test of $\eta = 1/4$

Finally we should like to make a remark about the large $N$-limit of the $O(N)$ model, which is often cited as evidence for the standard picture of $\beta_{crt} = \infty$ for all $N > 2$. As we pointed out in various places (see in particular [15]), this reasoning is faulty. It is true that at fixed $\tilde{\beta} \equiv \beta/N$ both $\xi$ and $\chi$ converge as $N \to \infty$ to the values of the spherical model, hence in that limit $\eta = 0$ up to log corrections. But, as we pointed out in [15], the approach to the limit is nonuniform, i.e. it becomes slower and slower as $\tilde{\beta} \to \infty$ or $\xi \to \infty$. With increasing $N$, $\ln \chi$ as a function of $\ln \xi$ will approach the spherical limit ($N = \infty$) form, but at any fixed $N$, for large enough $\beta$, we expect that it will fall away from that form and that $\ln \chi - 1.75 \ln \xi$ will aproach a constant for $\beta \to \infty$.

We are indebted to S.Caracciolo, R.G.Edwards, A.Pelissetto and A.D.Sokal for allowing us to use their data prior to publication; these data form the basis of the papers [13] and will apear in full in subsequent papers by [16].

# References

[1] J.M.Kosterlitz and D.J.Thouless, *J. Phys. (Paris)* **32** (1975) 581.

[2] J.Fröhlich and T.Spencer, *Commun.Math.Phys.* **81** (1981) 527.

[3] A.Patrascioiu, *Existence of Algebraic Decay in non-Abelian Ferromagnets*, University of Arizona preprint AZPH-TH/91-49.

[4] A.Patrascioiu and E.Seiler, *Percolation Theory and the Existence of a Soft Phase in 2D Spin Models*, Nucl.Phys.B.(Proc. Suppl.) **30** (1993) 184.

[5] A.Patrasciou and E.Seiler,*Phys.Rev.Lett.* **74** (1995) 1920.

[6] J.M.Kosterlitz, *J.Phys.* **C6** (1974) 1046.

[7] U.Wolff *Nucl.Phys.* **B334** (1990) 581.

[8] R.Gupta and C.F.Baillie, *Phys.Rev.B* **45** (1992) 2883.

[9] J.Apostolakis, C.F. Baillie and G.F.Fox, *Phys.Rev.D* **43** (1990) 2687.

[10] S.Caracciolo, R.G.Edwards, A.Pelissetto and A.D.Sokal, private communication.

[11] I.Affleck *Phys.Rev.Lett.* **66** (1991) 2429.

[12] W.Bietenholz, A.Pochinsky and U-J.Wiese, *Meron-Cluster Simulation of the $\theta$-Vacuum in 2-d O(3)-Model*, MIT preprint CPT 2433, hep-lat/9505019.

[13] S.Caracciolo, R.G.Edwards, A.Pelissetto and A.Sokal, *Phys.Rev.Lett.* **74** (1995) 2969;
– Nucl.Phys.B (Proc.Suppl.) **42** (1995) 752;
– *Asymptotic Scaling in the Two-Dimensional O(3) $\sigma$-Model at Correlation Length $10^5$*, hep-lat/9411009, to appear in *Phys. Rev. Lett.*

[14] A.Patrascioiu and E.Seiler, *Super-Instantons, Perfect Actions, Finite Size Scaling and the Continuum Limit*, preprint MPI-PhT/95-71, AZPH TH/95-17, hep-lat 9507018.

[15] A.Patrascioiu and E.Seiler, *Nucl.Phys.* **B 443** (1995) 596.

[16] S.Caracciolo, R.G.Edwards, A.Pelissetto and A.Sokal, in preparation.

**Tab.1a:** $\chi$ and $\xi$ for the $O(2)$ model on thermodynamic lattices ($L/\xi \geq 7$). Data are taken from [7].

| $\beta$ | $\xi$ | $\chi$ |
|---|---|---|
| .91 | 9.36(5) | 134.17(45) |
| .92 | 10.69(8) | 162.68(61) |
| .93 | 12.03(6) | 201.29(53) |
| .94 | 13.50(9) | 249.90(80) |
| .95 | 15.61(10) | 319.6(1.1) |
| .96 | 18.08(13) | 414.1(1.7) |
| .97 | 21.66(13) | 554.9(2.9) |
| .98 | 26.37(19) | 764.6(3.0) |
| .99 | 31.78(21) | 1092.(4.) |
| 1.00 | 40.20(20) | 1604.(4.) |
| 1.02 | 69.27(59) | 4170.(19.) |

**Tab.1b:** $\chi$ and $\xi$ for the $O(2)$ model on thermodynamic lattices ($L/\xi \geq 7$). Data are taken from [8].

| $\beta$ | $\xi$ | $\chi$ |
|---|---|---|
| .9090909 | 9.32(2) | 132.41(55) |
| .9345794 | 12.61(2) | 220.84(86) |
| .9478673 | 15.23(5) | 302.19(89) |
| .9615385 | 18.70(20) | 433.7(2.2) |
| .9708738 | 21.80(20) | 564.1(2.7) |
| .9756098 | 23.90(20) | 665.0(4.0) |
| .9803922 | 26.20(20) | 779.0(5.0) |
| .9900990 | 31.70(30) | 1087.(10.) |
| 1.000000 | 40.40(40) | 1631.(12.) |
| 1.010101 | 51.30(90) | 2520.(24.) |
| 1.020408 | 70.(1.) | 4258.(20.) |

**Tab.2a:** $\chi$ and $\xi$ for the $O(3)$ model on thermodynamic lattices ($L/\xi \geq 7$). Data are taken from [9].

| $\beta$ | $\xi$ | $\chi$ |
|---|---|---|
| 1.5 | 11.05(1) | 176.4(2) |
| 1.6 | 19.00(2) | 448.4(7) |
| 1.7 | 34.44(6) | 1263.7(3.3) |
| 1.75 | 47.2(2) | 2197.(15.) |
| 1.8 | 64.5(5) | 3823.(21.) |
| 1.85 | 88.7(5) | 6732.(25.) |
| 1.9 | 122.7(1.1) | 11867.(62.) |

**Tab.2a:** $\chi$ and $\xi$ for the $O(3)$ model on thermodynamic lattices ($L/\xi \geq 7$). Data are taken from [10].

| $\beta$ | $\xi$ | $\chi$ |
|---:|---:|---:|
| 1.5 | 11.13(16) | 178.0(1.1) |
| 1.55 | 14.69(15) | 281.5(1.7) |
| 1.6 | 18.66(35) | 447.6(2.6) |
| 1.65 | 25.21(13) | 742.8(2.0) |
| 1.675 | 29.77(13) | 971.9(2.7) |
| 1.7 | 34.66(13) | 1269.4(3.6) |
| 1.725 | 40.21(25) | 1661.8(5.8) |
| 1.75 | 47.20(19) | 2193.9(6.0) |
| 1.775 | 55.23(25) | 2886.4(10.2) |
| 1.8 | 65.17(57) | 3839.7(23.8) |